\AtBeginDocument{
    \newwrite\bibnotes
    \def\bibnotesext{Notes.bib}
    \immediate\openout\bibnotes=\jobname\bibnotesext
    \immediate\write\bibnotes{@CONTROL{REVTEX41Control}}
    \immediate\write\bibnotes{@CONTROL{apsrev41Control,author="08",editor="1",pages="1",title="0",year="1"}}
     \if@filesw
     \immediate\write\@auxout{\string\citation{apsrev41Control}}
    \fi
}
\documentclass[aps,pre,twocolumn,notitlepage,floatfix]{revtex4-1}
\usepackage{graphicx,amsmath,amssymb,verbatim,color,,xcolor,hyperref}
\hypersetup{breaklinks={true}}
\definecolor{green}{rgb}{0,0.7,0}

\begin{document}
\title{Critical behaviors of high-degree adaptive and collective-influence percolation}
\author{Jung-Ho Kim,$^*$ Soo-Jeong Kim,}
\thanks{Equal contribution.}
\address{Department of Physics, Korea University, Seoul 02841, Korea}
\author{K.-I. Goh}
\thanks{kgoh@korea.ac.kr}
\address{Department of Physics, Korea University, Seoul 02841, Korea}
\date{\today}

\begin{abstract}
How the giant component of a network disappears under attacking nodes or links addresses a key aspect of network robustness, which can be framed into percolation problems.
Various strategies to select the node to be deactivated have been studied in the literature; for instance, a simple random failure or high-degree adaptive (HDA) percolation.
Recently a new attack strategy based on a quantity called collective-influence (CI) has been proposed from the perspective of optimal percolation.
By successively deactivating the node having the largest CI-centrality value, it was shown to be able to dismantle a network more quickly and abruptly than many of the existing methods.  
In this paper, we focus on the critical behaviors of the percolation processes following degree-based attack and CI-based attack on random networks.
Through extensive Monte Carlo simulations assisted by numerical solutions, we estimate various critical exponents of the HDA percolation and those of the CI percolations.
Our results show that these attack-type percolation processes, despite displaying apparently more abrupt collapse, nevertheless exhibit standard mean-field critical behaviors at the percolation transition point. 
We further discover an extensive degeneracy in top-centrality nodes in both processes, which may provide a hint for understanding the observed results. 
\end{abstract}
\maketitle

{\bf
\noindent Network robustness is a longstanding problem in complex systems. 
Among various heuristics of optimal percolation proposed to effectively disintegrate a network, high-degree adaptive (HDA) attack and collective-influence (CI) attack are simple heuristics that use local information but can nevertheless destroy a network much more effectively than random percolation.
It is, however, largely unknown as yet how the critical behavior of percolation transition ({\em i.e.}, the manner in which the giant component disappears) is affected by various attack strategies.
In this work, it is revealed that despite apparently more abrupt disintegration the critical behavior of HDA and CI percolation transitions remains of standard mean-field type, accompanied by the massive degeneracy of top-centrality nodes near the critical point in these processes.
}

\section{Introduction}
\label{Sec:Introduction}

Statistical physics is the study of many-body problems.
Complex networks, in which many nodes interact via links, have been the subject of research by many statistical physicists \cite{2010NewmanNetworks}.
Understanding the robustness of complex networks is a longstanding problem \cite{2010CohenComplex}.
Observing how complex networks respond to intentional attacks is one common way to approach this problem.
The giant component of the network under intentional attack gets smaller and finally disappears at the critical point.
This can map to an optimal percolation problem that destroys the giant component of a network by deactivating a minimal set of nodes.
The minimal set of deactivated nodes is considered to be important nodes for maintaining the function of the network.
These important nodes are also called influencers \cite{2015MoroneInfluence}.

Unfortunately, finding the minimal set of nodes that can destroy the giant component in a network is in general an NP-hard problem.
Thus, various heuristic methods were developed for this problem.
In the early research stages, the nodes with high degree or high betweenness centrality were considered influencers \cite{2000AlbertError, 2000CallawayNetwork, 2001CohenBreakdown, 2002HolmeAttack}.
Since then, a number of strategies have been proposed to find near-optimal influencers more effectively \cite{2015MoroneInfluence, 2010KitsakIdentification, 2016BraunsteinNetwork, 2016ClusellaImmunization, 2016LuH, 2016MugishaIdentifying, 2016ZdeborovaFast, 2019RenGeneralized}.
These strategies are summarized in Ref.~\cite{2016LuVital}.
In general, each strategy assigns a certain centrality value to nodes in its own way, and the node with highest centrality value is deactivated one by one.
Centrality value of nodes can either be kept fixed as the initial value during the percolation process 
or be updated in every step ``adaptively'' as the nodes are deactivated. 

There is, however, no integrated perspective of various optimal percolation strategies yet.
An important step towards integrated understanding from the perspective of statistical physics is to characterize the percolation phase transition properties of each strategy.
It is well known that continuous mean-field phase transition occurs in the random percolation process on random networks \cite{2010NewmanNetworks,2010CohenComplex}.
On the other hand, several papers reported that the near-optimal percolation processes induce discontinuous collapse of the giant component at the transition point \cite{2016BraunsteinNetwork, 2016ClusellaImmunization, 2016MoroneCollective, 2016MugishaIdentifying, 2016ZdeborovaFast, 2017PeiEfficient, 2019RenGeneralized}.
Yet our understanding of how the universality of the percolation processes changes between these two extremes and what factors cause discontinuity is still limited.
Our primary aim in the current study is to contribute to make progress in this direction.

The percolation process we chose to study in this work about the critical behaviors of phase transition is the collective-influence (CI) percolation \cite{2015MoroneInfluence}.
CI percolation includes also the high-degree adaptive (HDA) percolation as a special case.
Principal reason why we chose CI percolation is that it is a near-optimal percolation process with abrupt yet continuous disintegration of the giant component at the critical point.
It is also known that CI percolation is able to attain near-optimal percolation results by straightforward and fast algorithm of time complexity $\mathcal{O}(N\log{N})$ \cite{2015MoroneInfluence, 2016MoroneCollective}.
 For this reason, it has been actively studied \cite{2017MoroneModel, 2018FerraroFinding} and refined \cite{2016MoroneCollective, 2017PeiEfficient}.
Detailed model description will be given in Sec. \ref{Sec:Model}.

\begin{figure}[t]
\centering
\includegraphics[width=\linewidth]{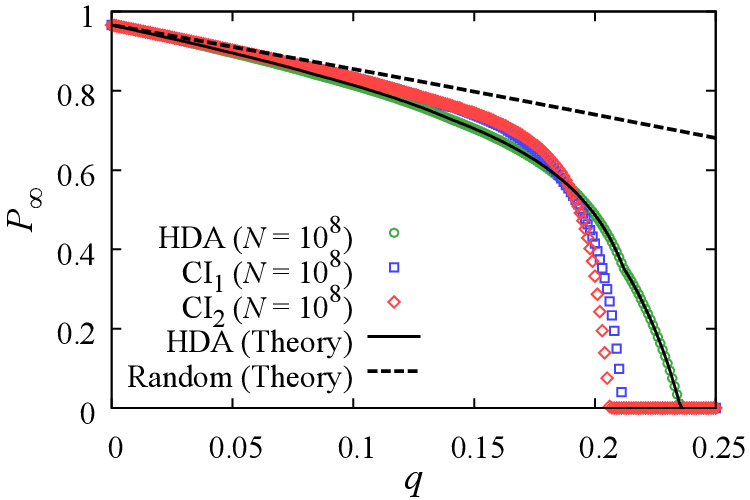}
\vspace*{-0.8cm}
\caption{
Plots of the order parameter, the probability $P_{\infty}$ that a randomly-chosen node belongs to the giant component, as a function of the deactivated-node fraction $q$ on ER networks with mean degree $z=7/2$ for various percolation processes defined in Sec. \ref{Sec:Model}.
Points are the results of Monte Carlo simulations with system size $N=10^{8}$ and averaged over more than 100 samples.
Solid line is for the numerical solution of HDA from Sec.~\ref{Sec:Method}, while the dashed line is for analytic solution of random percolation using the generating function method \cite{2000CallawayNetwork, 2001NewmanRandom}. 
}
\vspace*{-0.3cm}
\label{Fig:OrderParameter}
\end{figure}

As shown in the Fig. \ref{Fig:OrderParameter}, in the case of CI percolation, the order parameter seems to disappear rapidly in the vicinity of the critical point.
This raises the question if it might possess the order-parameter critical exponent $\beta$ different from that of random percolation.
To answer this question, we performed finite-size scaling analysis using extensive Monte Carlo simulations.
Despite the apparently abrupt change at critical point, we obtained standard mean-field critical exponents in both the HDA and CI percolations.
Moreover, to gain more insight of what is happening near the transition point, we examined the behavior of top-centrality nodes, discovering extensive degeneracy.

\section{Models}
\label{Sec:Model}
To select the influencers, we used the degree-centrality for HDA percolation and the CI-centrality for CI percolation \cite{2015MoroneInfluence}.
When a node is deactivated, the links attached to the node are also considered deactivated and in effect the degree of the deactivated node becomes zero.
Consequently, as the percolation process proceeds by successively deactivating nodes, a node's degree undergoes change. 
In the following, we denote node $i$'s degree, $k_{i}$, as the number of node $i$'s active neighbors in the remaining network.

In HDA percolation, one deactivates the node with highest degree and updates its neighbors' degree in every step. 
If there are more than one node with the same top-centrality (meaning largest degree), one of them is randomly deactivated.

For CI percolation, we follow the definition of CI-centrality in Ref.~\cite{2015MoroneInfluence}, given as follows:
\begin{equation}
\textrm{CI}_{\ell}(i)=(k_{i}-1) \sum \limits_{j \in \partial B (i, \ell)}^{} (k_{j}-1)~.
\end{equation}
Here $\textrm{CI}_{\ell}(i)$ is the CI-centrality value of node $i$ with distance parameter $\ell$, and $\partial B(i,\ell)$ is the set of nodes which are at distance $\ell$ (disallowing backtracking) from the node $i$.
The CI-centrality value of node $i$ incorporates not only its own degree but also the degrees of nodes which are at distance $\ell$ away from it.
Therefore, CI percolation does not simply deactivate high-degree nodes themselves, but tends rather to deactivate ``bridges'' between high-degree nodes \cite{2015MoroneInfluence}.
In CI percolation, one deactivates the node with top-centrality (meaning largest CI-centrality) and updates the CI-centrality values within the distance $(\ell+1)$ from the deactivated node each step.
Similarly to HDA percolation, if there are multiple nodes with same top-centrality, then a randomly-chosen node among them is deactivated. 
Note that by considering the distance-$0$ neighbor to be the node itself, we have  $\textrm{CI}_{0}(i)=(k_i-1)^2$. Thus HDA percolation can be treated as the special case of CI percolation with $\ell=0$ \cite{2015MoroneInfluence}.

\section{Monte Carlo simulation methods and numerical solutions}
\label{Sec:Method}
We applied HDA percolation and CI percolation with $\ell=1, 2$ on Erd\H{o}s-R\'enyi (ER) networks \cite{1960ErdosOn}.
The control parameter is the fraction of deactivated nodes $q$ ($0\le q\le1$). 
Primary quantity of interest is the probability $P_{\infty}$ that a randomly-chosen node among all nodes (regardless of remaining active or being deactivated) belongs to the giant component, which serves as the order parameter in the percolation process.
Numerically we calculate it by the largest connected component fraction averaged over Monte Carlo configurations. 
We also examine the average size $\chi$ of the component that a randomly-chosen node belongs to, which plays the role of susceptibility, defined as follows:
\begin{equation}
\chi(q)=\frac{\sum\limits_{\scriptsize{s,~\textrm{finite}}}^{}s^{2}n(s,q)}{\sum\limits_{\scriptsize{s,~\textrm{finite}}}^{}sn(s,q)}~,
\end{equation} 
where $s$ is the component size, $n(s,q)$ is the number of components with size $s$ at deactivated-node fraction $q$, and the summation runs over only finite sizes.
Numerically we calculate Eq.~(2) using the connected component configurations obtained by Monte Carlo simulations by omitting the largest connected component.
We have performed Monte Carlo simulations on ER networks with different mean degrees to find consistent results and in the following will present the results specifically for the mean degree $z=7/2$.

We also formulate the numerical solution for HDA percolation. For a random network, various quantities can be calculated by generating function method with degree distribution $p_{k}$  \cite{2000CallawayNetwork, 2001NewmanRandom}.
Fortunately, we can calculate degree distribution $p_{k}$ as a function of the deactivated-node fraction $q$ for HDA percolation using iteration method.
To apply the iteration method, we need the initial condition and the recurrence relations.
To begin with, the degree distribution of ER network of mean degree $z$ is well known to be the Poisson distribution $p_k={z^{k}e^{-z}}/{k!}$ \cite{1960ErdosOn}. 

To obtain the recurrence relations for the degree distribution as the nodes are deactivated, let the degree distribution $p_{k}(q)$ be the probability that the randomly chosen node has degree $k$ at the deactivated-node fraction $q$ in HDA percolation.
The degree distribution after additionally deactivating $dq$ fraction of nodes, $p_{k}(q+dq)$, can be obtained as follows.
First, as $q$ increases by $dq$, $p_{0}$ increases by $dq$ and $p_{K}$ decreases by $dq$ where $K=k_{\scriptsize{\textrm{max}}}(q)$, the maximum degree in the network at the deactivated-node fraction $q$. 
At the same time, when a node with degree $K$ is deactivated, the degree of the nodes linked to that deactivated node decreases by 1. 
The probability that a node with degree $k$ is connected to a link from the deactivated node is $kp_{k}/{\sum_{k'=0}^{\infty}{k'p_{k'}}}$. 
Combining the two effects, we obtain the complete set of recurrence relations as follows.
\begin{align}
\nonumber
p_{0}(q+dq)&= p_{0}(q)+K\frac{p_{1}}{\sum_{k'=1}^{\infty}{k'p_{k'}}}dq+dq~,\\ 
\nonumber
p_{1}(q+dq) &= p_{1}(q)+K\frac{2p_{2} - p_{1}}{\sum_{k'=1}^{\infty}{k'p_{k'}}}dq~,\\
&~~\vdots \\
\nonumber
p_{k}(q+dq)&= p_{k}(q)+K\frac{(k+1)p_{k+1} - kp_{k}}{\sum_{k'=1}^{\infty}{k'p_{k'}}}dq~,\\
\nonumber
&~~\vdots \\
\nonumber
p_{K}(q+dq) &= p_{K}(q)+K\frac{- Kp_{K}}{\sum_{k'=1}^{\infty}{k'p_{k'}}}dq-dq~.
\end{align}
The right hand side of each line contains a gain term due to the nodes with degree $(k+1)$ linked to the deactivated node and the loss term due to the nodes with degree $k$ linked to the deactivated node, except for the $k=0$ ($k=K$) equation that contains two gain (loss) terms. 
Theoretically, the maximum degree $K$ is not limited; however, when numerically solving Eq.~(3) we can practically introduce a finite maximum degree $K$ without sacrificing precision.
In our numerical calculation we set the cutoff probability to be $10^{-10}$, meaning that the degree $k$ with $p_k(q) <10^{-10}$ is considered nonexistent in the network.
In this way, the initial value of $K$ was obtained to be $K=21$ for the ER network with mean-degree $z=7/2$.
As Eq.~(3) is iterated, the maximum degree decreases when $p_K$ drops below the cutoff.  

Once the $p_k(q)$ is obtained by solving the recurrence relations Eq.~(3), one can apply the standard generating function technique \cite{2000CallawayNetwork, 2001NewmanRandom} to calculate $P_{\infty}(q)$ and $\chi(q)$ as a function of the deactivated-node fraction $q$. 
The underlying assumption behind this procedure is that after the HDA percolation, the remaining network can be characterized solely by $p_{k}(q)$ like random networks.
The agreement of the numerical solution with the Monte Carlo simulations, as shown in Fig.~\ref{Fig:OrderParameter}, validates this assumption.

\begin{figure*}[t]
\centering
\includegraphics[width=\linewidth]{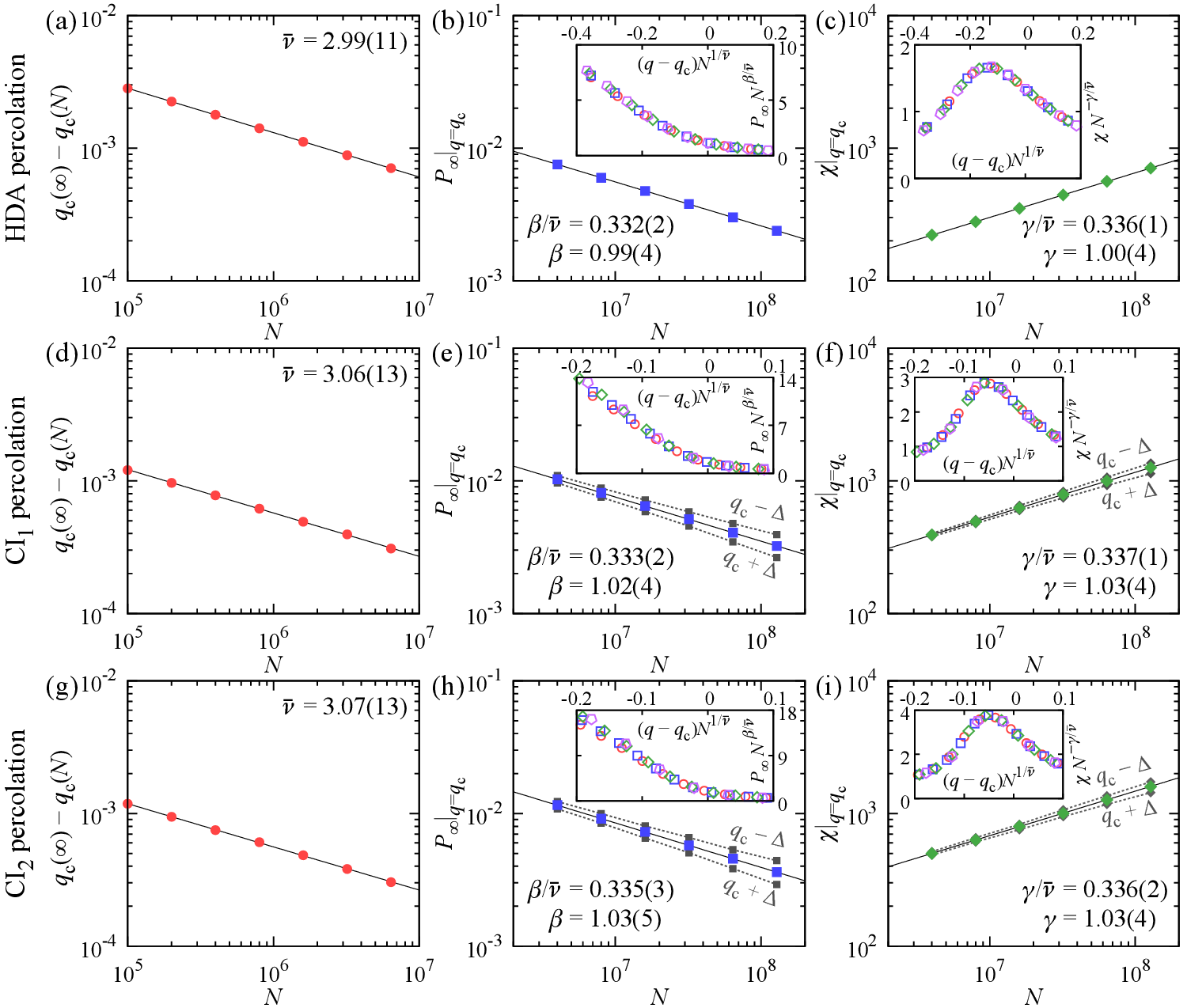}
\vspace*{-0.7cm}
\caption{
{\bf Finite-size scaling analysis results of HDA (a--c) and CI percolations (d--f for CI$_1$ and g--i for CI$_2$).}
Leftmost column (a,d,g) represents the relation of Eq. (\ref{Eq:FSSnu}), middle (b,e,h) and right (c,f,i) columns show the relation of Eqs. (\ref{Eq:FSSbeta}, \ref{Eq:FSSgamma}), respectively. 
Points represent the Monte Carlo simulation results and 
the solid black line is the fitting line at the critical point $q_c$. The gray dotted lines in (e,f,h,i) are the mere guidelines connecting the numerical simulation points (smaller-size points) obtained slightly away from the critical point $q_c$ by the amount $\Delta=0.000\,03$. 
We use the standard error for the error bars, which are smaller than the point size. Every fitted value has 95\% confidence interval. 
Each numerical simulation point was obtained by averaging over $3 \times 10^{3}$ to $6 \times 10^{6}$ samples, depending on the system size and the distance parameter $\ell$.
(Insets) Plots of finite-size-scaled data collapse curves from Eqs.~(\ref{Eq:DCbeta}, \ref{Eq:DCgamma}), using the obtained critical exponents. Simulation results with system sizes $N/10^6=1\rm~(circle),4\rm~(square),16\rm~(diamond),$ and $64\rm~(pentagon)$ are used for the inset plots.
}
\vspace*{-0.2cm}
\label{Fig:FSS}
\end{figure*}

\section{Critical behaviors}
\label{Sec:Critical}
As one deactivates nodes (hence increasing $q$), the giant component size decreases and becomes zero at the critical point $q_c$. 
Typical of continuous phase transitions, the average component size $\chi$ and the correlation length $\xi$ diverge at the critical point in the thermodynamic limit.
In the vicinity of critical point, the order parameter, the average component size and the correlation length are known to exhibit power laws with the critical exponent $\beta$, $\gamma$ and $\nu$, respectively: 
\begin{eqnarray}
P_{\infty}(q) &\propto (q_{c}-q)^{\beta} &~~~~~ \textrm{with}~q\rightarrow q_{c}^{-}~,\\
\chi(q) &\propto |q_{c}-q|^{-\gamma} &~~~~~ \textrm{with}~q\rightarrow q_{c}~,\\
\xi(q) &\propto |q_{c}-q|^{-\nu} &~~~~~ \textrm{with}~q\rightarrow q_{c}~.
\end{eqnarray}

We use finite-size scaling ansatz and Monte Carlo simulations to obtain these critical exponents.
According to the finite-size scaling theory, percolation quantities like $P_{\infty}$ and $\chi$ have the scaling form  near the critical point with the size of the system $N$ as
\begin{eqnarray}
\label{Eq:DCbeta}
P_{\infty}(N, q) = N^{-\beta/\bar{\nu}} \widetilde{P}[(q-q_{c})N^{1/\bar{\nu}}]~,\\
\label{Eq:DCgamma}
\chi(N, q) = N^{\gamma/\bar{\nu}} \widetilde{\chi}[(q-q_{c})N^{1/\bar{\nu}}]~,
\end{eqnarray}
where $\widetilde{P}$ and $\widetilde{\chi}$, respectively, is the scaling function. These equations imply the following power-law scalings,
\begin{eqnarray}
\label{Eq:FSSnu}
q_{c}-q_{c}(N) &\propto N^{-1/\bar\nu}~,\\
\label{Eq:FSSbeta}
P_{\infty}(N,q_c) &\propto N^{-\beta / \bar\nu} ~,\\
\label{Eq:FSSgamma}
\chi(N,q_c) &\propto N^{\gamma / \bar\nu}~, 
\end{eqnarray}
where $\bar\nu=d\nu$ with $d$ being effective dimension~\cite{2007HongFinite}, which is $d=6$ for ER network with random percolations.
In Eq.~(\ref{Eq:FSSnu}), $q_{c}(N)$ is the value of $q$ at which $\chi(q)$ displays maximum value in the networks of finite size $N$.

To apply the finite-size scaling theory, we first need to know the critical point $q_c$.
Using numerical solution from Sec.~\ref{Sec:Method}, the critical point of HDA percolation on ER network with mean degree $z=7/2$ is obtained to be $q_{c}=0.235\,550$.
For CI percolations, we estimated $q_{c}$ that best fits Eqs. (\ref{Eq:FSSbeta}, \ref{Eq:FSSgamma}) using Monte Carlo simulation results.
We note that using Eq.~(\ref{Eq:FSSnu}) for estimating $q_c$ was not as reliable due to relatively large errors in locating $q_c(N)$ from Monte Carlo simulations.
For the ER networks with the same mean degree $z=7/2$, we obtained the critical point of $\textrm{CI}_{1}$ percolation to be $q_{c}=0.211\,61(1)$, and for $\textrm{CI}_{2}$ percolation $q_{c}=0.206\,01(1)$.

\begin{table}[b]
\begin{center}
\begin{tabular}{lcccc}
\hline\hline
Model & $q_{c}$ & $\beta$ &$\gamma$ & $\bar\nu$ \\ 
\hline
HDA percolation & 0.235 550\phantom{()} & 0.99(4) & 1.00(4) & 2.99(11) \\
$\textrm{CI}_{1}$ percolation & 0.211 61(1) & 1.02(4) & 1.03(4) & 3.06(13) \\
$\textrm{CI}_{2}$ percolation & 0.206 01(1) & 1.03(5) & 1.03(4) & 3.07(13) \\
Random percolation & 0.714 285\phantom{()} & 1\phantom{.02(6)} & 1\phantom{.02(6)} & 3\phantom{.03(17)} \\
\hline\hline
\end{tabular}
\vspace*{-0.1cm}
\caption{The critical point and the critical exponents of various percolation processes on Erd\H{o}s-R\'enyi networks with mean degree $z=7/2$.}
\label{Tab:CPCE}
\end{center}
\end{table}

Fig.~\ref{Fig:FSS} displays, for the HDA, CI$_1$, and CI$_2$ percolations, the finite-size scaling results of Eqs.~(\ref{Eq:FSSnu}--\ref{Eq:FSSgamma}) in the main panels, together with corresponding finite-size-scaled data collapse curves of Eqs.~(\ref{Eq:DCbeta}--\ref{Eq:DCgamma}) in the insets. 
This analysis provides estimates of the exponents $\bar{\nu}$, $\beta/\bar{\nu}$, and $\gamma/\bar{\nu}$, from which we extract values of the critical exponents $\beta$, $\gamma$, and $\bar{\nu}$. 
The obtained values of the critical point and the critical exponents are summarized in Table~\ref{Tab:CPCE}.
The critical exponents we obtained are consistent with those of random percolation within margin of error. 
Our results show that, despite the apparently more rapid change near the transition point, both HDA and CI percolations still belong to the same universality as random percolation in random networks. 

\begin{figure*}[t]
\centering
\includegraphics[width=\linewidth]{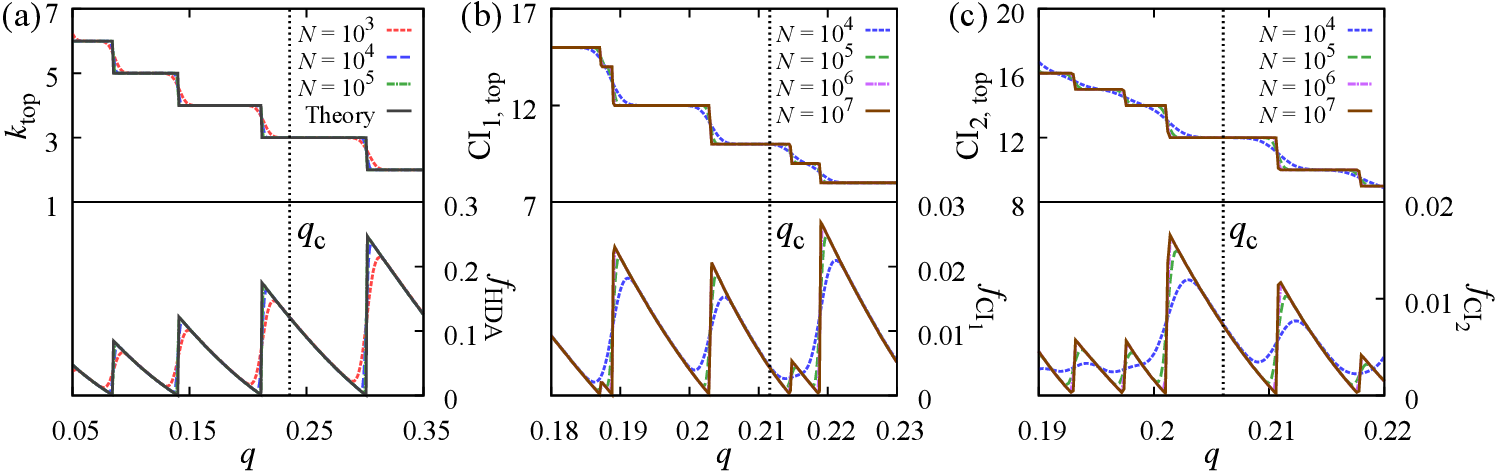}
\vspace*{-0.7cm}
\caption{
{\bf Properties of the top-centrality value during the HDA (a) and the CI (b, c) percolations.}
(Upper panels, left tics) Plots of the top-centrality value, $k_{\rm top}$ or ${\rm CI}_{\ell,\rm top}$, as a function of the deactivated-node fraction $q$.
(Lower panels, right tics) 
Plots of the fraction $f$ of the nodes (with respect to $N$) that have the top-centrality value as a function of the deactivated-node fraction $q$.
Both plots contain data from simulations averaged over $10^{3}$ to $10^{6}$ samples.
Vertical guideline indicates the location of the critical deactivated-node fraction $q_c$ at which the giant component disappears.
Black solid lines in (a) are obtained from the numerical solutions of HDA from Sec.~\ref{Sec:Method}.
}
\vspace*{-0.2cm}
\label{Fig:Degeneracy}
\end{figure*}

\begin{figure}[t]
\centering
\includegraphics[width=\linewidth]{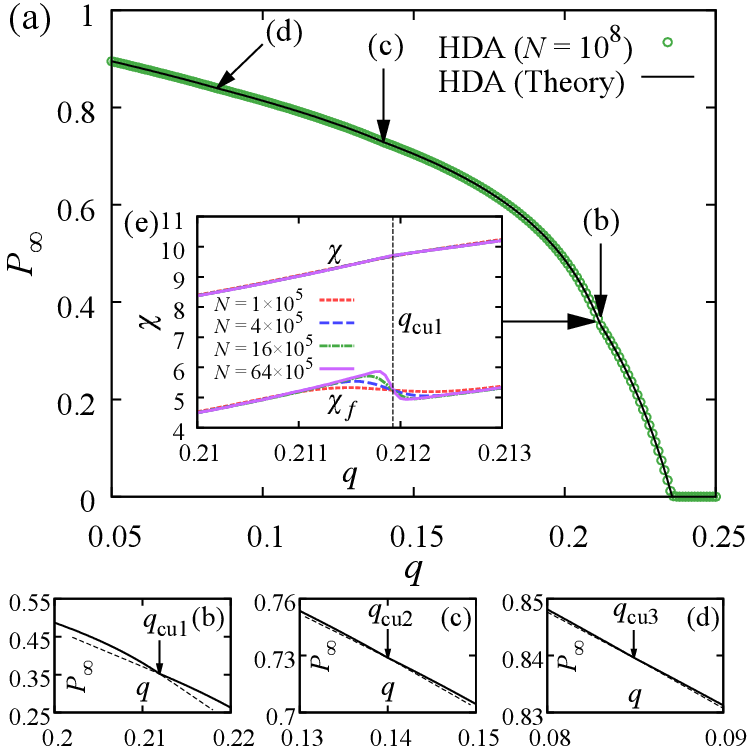}
\vspace*{-0.7cm}
\caption{
{\bf (a)} Plot of the order parameter $P_{\infty}$ of HDA percolation on ER networks with mean degree $z=7/2$ as a function of the deactivated-node fraction $q$.
Points are results of Monte Carlo simulations with networks of size $N=10^{8}$, averaged over more than 100 samples.
Black solid lines are numerical solutions from Sec.~\ref{Sec:Method}. 
{\bf (b--d)} The close-up curve near the cusp point, (b) for $q_{\rm cu1}=0.211\,925$, (c) for $q_{\rm cu2}=0.140\,014$, and (d) for $q_{\rm cu3}=0.084\,917$.
Dashed lines in (b--d) are the tangential line at each side of the cusp point, illustrating its non-analyticity.
{\bf (e)} Behavior of the average component size $\chi$ (upper) and the fluctuation of the order parameter $\chi_f$ (lower) near the cusp point at $q_{\rm cu1}=0.211\,925$ as a function of the deactivated-node fraction $q$.
Data are from Monte Carlo simulations with various system size $N=1,4,16,64 \times 10^{5}$, averaged over $3 \times 10^{4}$ to $1 \times 10^{6}$ samples.
Here, the fluctuation of the order parameter $\chi_f(q)$ is defined by $\chi_f(q)=N(\langle P_{\infty}(q)^2\rangle - \langle P_{\infty}(q)\rangle^2)$, where $\langle \cdots\rangle$ denotes the ensemble average.
}
\vspace*{-0.2cm}
\label{Fig:Cusp}
\end{figure}

\section{Degeneracy of top-centrality value}

Despite the apparently more abrupt collapse, the results of our analysis strongly suggest that HDA and CI percolations belong to the same universality as random percolation.
To gain more insight, we examine how the top-centrality value (degree-centrality for HDA percolation and CI-centrality value for CI percolation) behaves near the percolation critical point (Fig.~\ref{Fig:Degeneracy}).
First we observe that as the percolation process proceeds and approaches to the critical point, the top-centrality value becomes highly degenerate, which is realized as the succession of long plateaus in the upper panels of Fig.~\ref{Fig:Degeneracy}. 
For HDA, the lines for different system size $N$ converge to the stairs-like curve predicted by the numerical solution as the system size increases. Same limiting behavior is observed for the CI$_{1,2}$ percolations, suggesting that they would develop the stairs-like behavior asymptotically as well.
Next, we show the fraction of top-centrality nodes in the lower panels of Fig.~\ref{Fig:Degeneracy}.
It also tends to the limiting saw-like curve as the system size increases, implying that the number of top-centrality nodes near the critical point $q_{c}$ is extensive, that is, proportional to the system size $N$. 
In other words, if the system size goes to infinity, the number of nodes with the top-centrality value would also go to infinity, confirming a massive degeneracy of top-centrality value.

According to the rules of HDA and CI percolation, when there are many nodes with the same top-centrality value, one of the nodes is randomly selected and deactivated.
The presence of extensive degeneracy of top-centrality nodes near the critical point implies that one of the extensively-many top-centrality nodes is chosen randomly and deactivated.
This induces that the selectiveness is weakened and the node-deactivation process would become more random-like.

The presence of extensive degeneracy in top-centrality value manifests itself in the behavior of order parameter as well. 
In HDA percolation, the cusp points in the order parameter curve are observed at the points where $k_{\rm top}$ change its value (see for example the cusps at $q_{\rm cu1}=0.211\,925$, $q_{\rm cu2}=0.140\,014$, and $q_{\rm cu3}=0.084\,917$ in Fig.~\ref{Fig:Cusp}). 
The discrete change of $k_{\rm top}$ value across the cusp point induces discontinuous change of the rate of decrease of the order parameter across that point, establishing the cusp [see Fig.~\ref{Fig:Cusp}(b--d)].
This phenomenon is also verified by the Monte Carlo simulation near $q_{\rm cu1}$.
The susceptibility-like quantities, the average component size $\chi$ and the fluctuation of the order parameter $\chi_f$, also exhibit non-analytic behavior such as cusp ($\chi$) or discontinuity ($\chi_f$) at the cusp point, as shown in Fig.~\ref{Fig:Cusp}(e). Here $\chi_f$ is defined by $\chi_f(q)=N(\langle P_{\infty}(q)^2\rangle - \langle P_{\infty}(q)\rangle^2)$ where $\langle \cdots\rangle$ denotes the ensemble average.
They do not, however, diverge there. These cusp points, therefore, are not likely critical points.
The existence (or absence) of similar cusps in the CI percolations is not clearly discernible due to limited resolution of our Monte Carlo simulations.

\section{Discussion}
In this paper, we have studied the critical behaviors of HDA and CI percolation transitions on random networks using Monte Carlo simulations and numerical solutions.
We found that the critical behavior of the two attack-based processes is in the same universality class as the random percolation, despite their more abrupt, near-optimal disruption of network connectivity than random percolation.
We uncovered the massive degeneracy of maximum centrality value near the critical point, which might contribute to render the transition of mean-field type.
Recently, a study reported \cite{2020AlmeiraScaling} non-standard-mean-field critical exponent for HDA percolation, specifically $\bar\nu=2.59(7)$, which contradicts with our mean-field-consistent result $\bar\nu=2.99(11)$ in this work. 
Our analysis uses much larger network size (up to ${\cal O}(10^8)$ compared to $6.4\times10^4$ in Ref.~\cite{2020AlmeiraScaling}) and is supported also by the numerical solutions.
It is noteworthy that Ref.~\cite{2020AlmeiraScaling} also reported high-betweenness adaptive percolation to exhibit different critical exponents from those of random percolation.
It has also been reported that HDA percolation has the same exponents as random percolation on two-dimensional proximity graphs \cite{2006NorrenbrockFragmentation}.
This also indirectly supports that HDA percolation belongs to the same universality class as random percolation.

In a broader perspective, our work initiates the studies of attack-type and optimal percolation processes for deeper understanding from the viewpoint of critical behaviors at the percolation transition point.
Understanding the nature of critical behaviors will help for example devise detection and protection strategy upon the attacks on the network, as different criticality entails different `early warnings' \cite{2009EarlyScheffer}.
Another interesting standing question is the condition for the non-mean-field criticality in the optimal percolation processes.
The HDA and CI percolation processes implement local heuristics for selecting nodes to be deactivated.
There exist approaches based on global optimization methods \cite{2016BraunsteinNetwork, 2016ClusellaImmunization, 2016MoroneCollective, 2016MugishaIdentifying, 2016ZdeborovaFast, 2017PeiEfficient, 2019RenGeneralized}. that reportedly produce discontinuous disappearance of the giant component in the attacked network.
Identifying minimal condition for the non-standard-mean-field criticality as well as the discontinuity in the network attack process remains a theoretically intriguing problem, to be explored in future works.
Finally, the phase transition property of the CI percolation with higher-order network effects such as on modular networks is also an interesting problem for future study.

\section*{Acknowledgments}
This work was supported in part by the National Research Foundation of Korea (NRF) grants funded by the Korea government (MSIT) (No.\ 2017R1A2B2003121 and No.\ 2020R1A2C2003669).

\section*{Author contributions}
J-HK, S-JK and K-IG conceived the project; J-HK, S-JK performed the simulations; all authors interpreted the results and contributed to the writing of the manuscript.
\bibliography{CI_criticality_R2}
\end{document}